# Primordial gravitational waves and the $H_0$-tension problem

Leila L. Graef,[1,6,*] Micol Benetti,[2,3,4,†] and Jailson S. Alcaniz[2,5,‡]

[1]*Instituto de Física, Universidade Federal Fluminense, Avenida General Milton Tavares de Souza s/n, Gragoatá, 24210-346 Niterói, Rio de Janeiro, Brazil*
[2]*Observatório Nacional, Rua General José Cristino 77, 20921-400, Rio de Janeiro, Rio de Janeiro, Brazil*
[3]*Physics department "E. Pancini," Napoli University "Federico II," Via Cinthia, I-80126, Napoli, Italy*
[4]*Istituto Nazionale di Fisica Nucleare (INFN), Napoli section, Via Cinthia 9, I-80126, Napoli, Italy*
[5]*Departamento de Física, Universidade Federal do Rio Grande do Norte, 59072-970, Natal, Rio Grande do Norte, Brazil*
[6]*Universidade do Estado do Rio de Janeiro, Instituto de Física – Departamento de Física Teórica – Rua São Francisco Xavier 524, 20550-013, Maracanã, Rio de Janeiro, Brazil.*

We analyze the $H_0$-tension problem in the context of models of the early universe that predict a blue tilted spectrum of primordial gravitational waves (GWs), which is a positive value of the tensor tilt $n_T$. By considering the GW's contribution, $N_{eff}^{GW}$, to the effective number of relativistic degrees of freedom, $N_{eff}$, and assuming standard particle physics, we discuss the effects of $N_{eff}^{GW}$ on the background expansion, especially the constraints on the Hubble parameter $H_0$. We analyze three scenarios that take into account the contribution of $N_{eff}^{GW}$ using recent data of cosmic microwave background, baryon acoustic oscillation, the latest measurement of the local expansion rate, along with the LIGO constraints on the tensor to scalar ratio, $r$, and the tensor index. For the models explored, we show that an additional contribution from the primordial GW's background to $N_{eff}$ does not solve but alleviates the current $H_0$-tension problem.

## I. INTRODUCTION

The search for gravitational waves (GWs) is one of the main goals of modern cosmology. Along with current cosmic microwave background (CMB) data, the results from experiments such as LIGO [1–3] and VIRGO [4,5] are able to provide constraints not only on models of the primordial universe, imposing limits on the tensor to scalar ratio, $r$, and on the tensor spectral index, $n_t$ [6], but also on the physics of the late-time cosmic acceleration (see, e.g., [7,8]). Currently, some of these experiments have allowed one to test the viability of several models of the early universe, and the present constraints are compatible with both the simplest slow-roll scenarios of inflation [9] and some alternative scenarios [10–15].

In this context, it is worth mentioning that the recent CMB polarization data have excluded a much larger range of red tensor tilt values (negative values of the tensor spectral index) than of blue ones (positive values of the tensor spectral index) [6,16,17]. A possible detection of a blue tensor tilt in the future would rule out a large class of models of inflation, since the so-called consistency relation would not be fulfilled in the standard scenario without violating the null energy condition. Future experiments, such as the Hanford-Livingston Virgo (HLV) network [18] (and also the Hanford-Livingston Virgo-Japan-India (HLVJI) network), the future space-based detector DECIGO [19], satellite missions such as LiteBIRD [20], and ground based experiments such as CMB-S4 [21], are expected to give great improvement to the current limits on primordial GWs (see also [22–25]).

The effects of the GW's background on the CMB and matter power spectra are similar to those of massless neutrinos, unless the initial density-perturbation amplitude for the GW's gas is nonadiabatic [26,27]. The primordial GWs also affect the expansion of the universe. Since they are relativistic degrees of freedom, they add to the effective number of relativistic species, $N_{eff} + N_{eff}^{GW}$ [28,29], which increases the radiation energy density and decreases the redshift of matter-radiation equality, according to the relation $1 + z_{eq} = \Omega_m/\Omega_r = \Omega_m/\Omega_\gamma(1 + 0.2271N_{eff})$. In standard inflationary models, which predict a small red tensor tilt, the contribution from $N_{eff}^{GW}$ is negligible. However, the same is not generally true for models that predict a primordial spectrum with a blue tensor tilt (see, e.g., [13–15,17,30]).[1]

Another important aspect worth considering is that a higher contribution of $N_{eff}$ in the early universe leads to a faster expansion and hence a smaller value of the sound horizon at recombination, $r_*$, and a consequent higher value of the current expansion rate, $H_0$ [33]. For this reason, higher values of $N_{eff}$ have been considered as an attempt to alleviate the so-called $H_0$-tension problem [34–36], i.e., the discrepancy on the Hubble parameter value predicted in the context of the $\Lambda$CDM model using the CMB data, i.e., $H_0 = 66.93 \pm 0.62$ kms$^{-1}$ Mpc$^{-1}$ [37], and the value obtained from different geometric distance calibrations of Cepheids using observations of the Hubble Space Telescope (HST), $H_0 = 73.52 \pm 1.62$ kms$^{-1}$ Mpc$^{-1}$ [38,39] (see also [40] and references therein for an overview of the discussion). While previous data from the Planck Collaboration prefer larger values of $N_{eff}$ [34,35], the latest data released [41] show a preference for smaller values. Despite that, the current data still allow values of $N_{eff}$ higher than the one predicted by the standard model ($N_{eff} = 3.046$). Although extra neutrino species are not predicted by the standard particle physics, primordial GWs with a blue tensor spectrum can naturally provide a significant contribution for $N_{eff}$ and, therefore, may alleviate the current $H_0$-tension problem.

*leilagraef@if.uff.br
†micolbenetti@on.br
‡alcaniz@on.br

[1]The contribution of $N_{eff}^{GW}$ to the radiation content of the universe also affects the predictions of the primordial nucleosynthesis (BBN) [31,32].

A careful analysis of some possible effects of a large positive tensor tilt on cosmological parameters was performed in Refs. [6,26]. Starting from the results obtained in these works, here we focus specifically on the impact of $N_{eff}^{GW}$ on the constraints on the parameter $H_0$, in light of the recent Planck Planck + HST results, analyzing its consequences to the problem of the tension in the $H_0$ measurements. We also analyze the sensitivity of the results with respect to the choice of the ultraviolet (UV) frequency cutoff, $k_{UV}$, clarifying how much this choice can affect the results obtained here and in the previous works mentioned. We concentrate on models that predict a primordial tensor spectrum with a blue tilt, since in this case a significant contribution of $N_{eff}^{GW}$ is expected. We organized this paper as follows. In Sec. II we review the theoretical framework and discuss the basic assumptions considered. The method, observational dataset, and priors used in the computational analysis are discussed in Sec. III. In Sec. IV we present our results, and in Sec. V we summarize our conclusions.

## II. THEORY

The spectrum of cosmological perturbations is a fundamental quantity to connect the physics of the early universe with observations. The primordial GWs are described by the dimensionless tensor power spectrum $P_t(k)$,

$$P_t(k) = A_t(k_*) \left( \frac{k}{k_*} \right)^{n_t(k_*)}, \qquad (1)$$

where the spectral index $n_t$ parametrizes the dependence of $P_t(k)$ on the comoving wave number $k$. This follows in analogy to the standard scalar spectrum parametrization, in which $A_s$ and $n_s$ describe the scalar amplitude and the scalar spectral index, respectively, and the tensor to scalar ratio, $r = A_t/A_s$, is the quantity that relates the amplitude of both the tensor and the scalar spectrum. The consistency relation of standard inflation, $r = -8n_t$, implies that $n_t$ is always negative in these models. However, the same is not true in the case of other early universe scenarios [13–15,17,30].

The primordial tensor spectrum can also be written as a function of the frequency $f$ in the following form:

$$P_t(f) = rA_s \left( \frac{f}{f_*} \right)^{n_t} = rA_s \left( \frac{f/Hz}{1.6 \times 10^{-17}} \right)^{n_t} \qquad (2)$$

for a pivot scale $k_* = 0.01$ Mpc$^{-1}$ and $f/Hz = 1.6 \times 10^{-15}$ k/Mpc$^{-1}$. The energy density of gravitational waves is given by

$$\rho_{GW} = \int_0^{k_{UV}} d\log k \frac{P_t(k)}{32\pi Ga^2} [\mathcal{T}'(k,\eta)]^2, \qquad (3)$$

where $k_{UV}$ is an ultraviolet cutoff whose values will be discussed later and $\mathcal{T}(k,\eta)$ is the tensor transfer function. By performing this integral using the conformal time derivative of transfer function during the radiation domination, it is possible to show that [26]

$$\begin{aligned} \rho_{GW} &= \frac{A_s r}{32\pi G} \left( \frac{k_{UV}}{k_*} \right)^{n_t} \frac{1}{2n_t(a\eta)^2} \\ &= \frac{A_s r}{24 n_t} \left( \frac{k_{UV}}{k_*} \right)^{n_t} \rho_{tot}, \end{aligned} \qquad (4)$$

where in the second equality we considered that during this period $1/(a\eta)^2 = H^2 = 8\pi G\rho_{tot}/3$. At this epoch the total energy density is given by the sum of photons, neutrinos, and GW energy densities, $\rho_{tot} = \rho_\gamma + \rho_\nu + \rho_{GW}$, which can be rewritten as

$$\rho_{tot} = \rho_\gamma \left( 1 + \frac{7}{8} \left( \frac{4}{11} \right)^{4/3} N_{eff} \right). \qquad (5)$$

By substituting the above expression in the right-hand side of Eq. (4) we obtain

$$\begin{aligned} \rho_{GW} &= \frac{A_s r}{24 n_t} \left( \frac{k_{UV}}{k_*} \right)^{n_t} \left( 1 + \frac{7}{8} \left( \frac{4}{11} \right)^{4/3} N_{eff} \right) \rho_\gamma \\ &= \frac{7}{8} \left( \frac{4}{11} \right)^{4/3} (N_{eff} - 3.046) \rho_\gamma. \end{aligned} \qquad (6)$$

In our analysis we assume that beyond photons and three families of standard model neutrinos, the radiation energy density is also made up of gravitational waves, with no extrarelativistic degrees of freedom, i.e., $N_{eff} = 3.046 + N_{eff}^{GW}$. Solving for $N_{eff}$ we find

$$N_{\text{eff}} = \frac{\frac{8}{7}\left(\frac{11}{4}\right)^{4/3}\left[\frac{A_s r}{24 n_t}\left(\frac{k_{\text{UV}}}{k_*}\right)^{n_t}\right] + 3.046}{1 - \left[\frac{A_s r}{24 n_t}\left(\frac{k_{\text{UV}}}{k_*}\right)^{n_t}\right]}. \tag{7}$$

The above equations should only be trusted in the regime $\rho_{\text{GW}}/\rho_{\text{tot}} \ll 1$, which is the case considering the current observational constraints. We assume in our analysis that the power law form of the tensor power spectrum holds all the way up to the cutoff frequency $k_{\text{UV}}$—see [42] for a discussion on different approaches. We also assume the standard thermal history with an instantaneous transition from the phase of the early universe responsible for producing the primordial spectrum to the phase of radiation domination.

By neglecting the contribution of gravitational waves to $\rho_{\text{tot}}$ in Eq. (4) one obtains the following approximation[2]:

$$N_{\text{eff}} = 3.046 + \left[3.046 + \frac{8}{7}\left(\frac{11}{4}\right)^{4/3}\right]\frac{A_s r}{24 n_t}\left(\frac{k_{\text{UV}}}{k_*}\right)^{n_t}. \tag{8}$$

The above relation is illustrated in Fig. 1 for some selected values of the parameter $n_t$ with $r = 10^{-3}$ and $A_s = 2.2 \times 10^{-9}$. Clearly, there is a steep increase of $N_{\text{eff}}$ as a function of $k_{\text{UV}}$ showing that the bigger the value of $n_T$, the smaller the value of $k_{\text{UV}}$ at which this steep increase occurs.

Going back to Eq. (3), one needs to impose both IR and UV cutoffs to the integral in this equation. The only modes that contribute to the radiation energy density are those inside the horizon at a given time since those modes oscillate, propagating as massless modes. This implies that the IR cutoff is a time-dependent quantity. As a consequence, the contribution of $N_{\text{eff}}^{\text{GW}}$ that affects big bang nucleosynthesis (BBN) is different from the contribution of $N_{\text{eff}}^{\text{GW}}$ that affects CMB, e.g., However, in the case of a blue tilted tensor spectrum the IR contribution to $\rho_{\text{GW}}$ is negligible compared to the UV contribution.[3]

The most interesting cases for us here correspond to the cases of blue tensor spectra since we are interested in analyzing the contribution of $N_{\text{eff}}^{\text{GW}}$, and such a contribution is negligible in the case of red tensor tilt. Despite that, we also extend some analysis to red tilts for completeness. We also consider that $N_{\text{eff}}^{\text{GW}}$ does not depend on time for a fixed UV cutoff.

---

[2]This is equivalent to the first order expansion of Eq. (7) around $\rho_{\text{GW}}/\rho_{\text{tot}} = 0$.

[3]The effect of considering the correct infrared cutoff would imply in additional term in Eq. (8) such that its second line becomes $(A_s r/24 n_t)[(k_{\text{UV}}/k_*)^{n_t} - (k_{\text{IR}}/k_*)^{n_t}]$. The smaller value considered for $k_{\text{UV}}/k_*$ implies $(k_{\text{UV}}/k_*)^{n_t} = 10^{24 n_t}$. We can consider $k_{\text{IR}}$ as the horizon size at decoupling $k \approx 1 \text{ Mpc}^{-1}$ [6], which leads to $(k_{\text{IR}}/k_*)^{n_t} = 63^{n_t}$. For any positive value of $n_t$ this term is completely negligible compared to the one coming from the UV contribution. On the other hand, when $n_t$ is negative, the total second term in the equation above is always orders of magnitude smaller than the first one (3.046) for all the allowed values of $n_t$ ($n_t > -0.6$). Therefore the resulting $N_{\text{eff}}^{\text{GW}}$ is too small to have a significant contribution. In conclusion, neglecting the IR cutoff, as it will be considered in what follows, does not affect any of the analysis performed in this work.

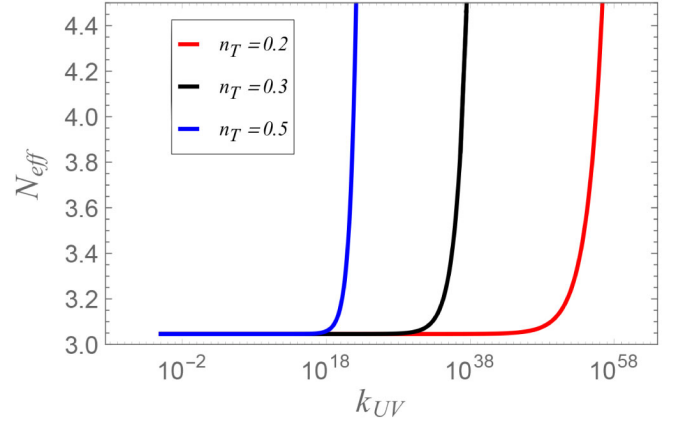

FIG. 1. $N_{\text{eff}}$ as a function of $k_{\text{UV}}$ for some particular values of $n_T$.

For the UV cutoff several possibilities have been considered in the literature (see, e.g., [6,26,31]). First, we perform a more general test, without assuming any specific value for the cutoff, in order to see how the results are sensitive to this choice. Looking at Fig. 2 we note that lower values of $k_{\text{UV}}$ allow higher values for $n_t$ and $r$.

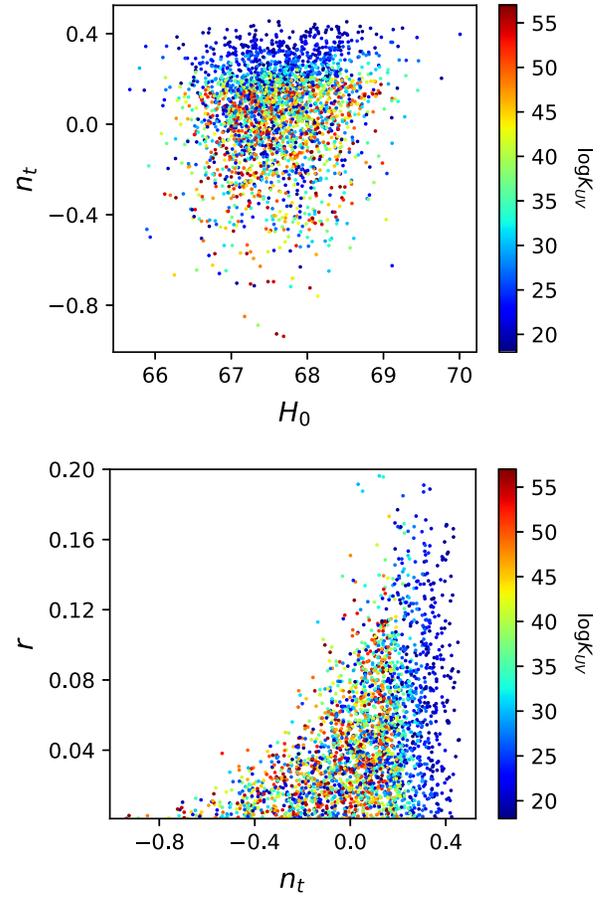

FIG. 2. A scatter plot showing the relation among $n_T$, $H_0[\text{km/s/M}_{\text{pc}}]$, and $k_{\text{UV}}[\text{M}_{\text{pc}}^{-1}]$ and the relation among $n_T$, $r$, and $k_{\text{UV}}$ for the base dataset.

Because of this behavior, we choose to consider two very different values for the cutoff, which are both model independent. In the first one (hereafter Model 1), inspired by the approach of [31], we simply take it to be given by the grand unification theory scale, i.e., $10^{16}$ GeV. This corresponds to consider the ratio $k_{UV}/k_* \approx 10^{56}$ (or equivalently, $f_{UV} \approx 10^{40}$ Hz). In a second case (hereafter Model 2), we consider the ratio $k_{UV}/k_* \approx 10^{24}$ ($f_{UV} \approx 10^8$ Hz), as suggested in [26]. This latter choice follows from the assumption that the power law spectrum extends over $\approx 60$ $e$-folds and is chosen not due to some particular mechanism for generating the gravitational waves in the early universe but instead to correspond to the expected amount of expansion. This value, nevertheless, coincides approximately with the value chosen in [6] from inflationary arguments.

## III. ANALYSIS

After having performed a preliminary analysis with free $k_{UV}$, shown in Fig. 2, in what follows we work with the two values selected above for $k_{UV}$ in order to verify how sensitive the results are to this difference on the cutoff. Therefore, we consider two cases of the $\Lambda CDM + r + n_t$ model with $k_{UV}/k_* \approx 10^{56}$ (Model 1) and $k_{UV}/k_* \approx 10^{24}$ (Model 2). In both the contribution $N_{eff}^{GW}$ is treated as a derived parameter that is a function of $r$ and $n_t$. This model is parametrized by the free parameters $r$ and $n_t$ in addition to the usual set of cosmological parameters: the baryon density, $\Omega_b h^2$; the cold dark matter density, $\Omega_c h^2$; the ratio between the sound horizon and the angular diameter distance at decoupling, $\theta$; the optical depth, $\tau$; the primordial scalar amplitude, $A_s$; and the primordial spectral index $n_s$. For the sake of comparison we also analyze, as a reference, the model $\Lambda CDM + r + n_t$, which does not consider the contribution of GWs to $N_{eff}$. In all cases we analyze the constraints on the derived parameter $H_0$. We work with flat priors and purely adiabatic initial conditions, also fixing the sum of neutrino masses to 0.06 eV. In the preliminary analyses made with free $k_{UV}$, we worked with a flat prior on the logarithm of $k_{UV}$ in the range from 18 to 57, therefore covering the range from $k_{UV} = 10^{18}$ Mpc$^{-1}$ to $k_{UV} = 10^{57}$ Mpc$^{-1}$.

We perform cosmological analysis considering two datasets. The first, called "base," includes the CMB dataset from the recent release of the Planck Collaboration [37] (hereafter PLC), considering both the high-$\ell$ and low-l Planck temperature data[4] and the $B$-mode polarization data from the BICEP2 Collaboration [44,45] to constrain the parameters associated with the tensor spectrum, using the combined BICEP2/Keck-Planck likelihood (hereafter BKP). Also, such a dataset considers a Gaussian prior on the optical depth parameter $\tau = 0.055 \pm 0.009$ recovered from Planck high frequency instrument large angular scale polarization measurements [41,46] joined with baryon acoustic oscillation (BAO) data from the 6dF Galaxy Survey (6dFGS) [47], Sloan Digital Sky Survey (SDSS) DR7 Main Galaxy Sample galaxies [48], BOSS galaxy samples, LOWZ and CMASS [49]. Finally, we also include data from GW direct detection experiments. The ground-based interferometers can probe the primordial gravitational wave spectrum in a range of frequencies from 1 Hz to $10^4$ Hz, providing us with upper limits on the parameters $r$ and $n_t$. Here, in particular, we use data from LIGO [1–3]. The information from LIGO is implemented as a Gaussian prior on $n_t$. We note that pulsars also can be used to constrain the tensor parameters. However, as shown in [6] the upper limits provided by LIGO are more constraining than the ones provided by pulsars. For this reason we do not include data from pulsars in our analysis.

Therefore, summing up our "base" dataset combines PLC + BKP + BAO + LIGO data, while the second dataset we assume also considers the results of Riess $et$ $al.$ on the local expansion rate, $H_0 = 73.52 \pm 1.62$ kms$^{-1}$ Mpc$^{-1}$ (68% C.L.), based on direct measurements made with the Hubble Space Telescope and Gaia [39]. Hereafter we denote this second dataset as "base + HST."

In our analysis we use the more recent release of the package CosmoMC [50], assuming scalar and tensor pivot scale $k_* = 0.01$ Mpc$^{-1}$, since at this value the BKP release data are most sensitive and it is close to the decorrelation scale between the tensor amplitude and slope for Planck and BKP joint constraints [16].

## IV. RESULTS AND DISCUSSION

The main results of our analysis are shown in Fig. 3, where confidence regions in the plane $n_T$ vs $H_0$ for the two cutoff choices are displayed. One can see that the Model 2 (red), which has the smaller cutoff value, allows higher values of $n_T$ than Model 1 (blue). Also, in the case of the standard model without a consistency relation[5]—the reference model in gray—it allows even higher values for $n_T$. This is due to the fact that by considering the contribution of GWs to $N_{eff}$, the upper limits imposed on this quantity by the data restrict the allowed values of $n_T$. Another important point worth noting is that for both models, higher values of $H_0$ are allowed for higher values of $n_T$. The allowed range of $H_0$ is even more pronounced for the dataset base + HST, which is shown in the right graph of Fig. 3, where one can observe a steep increase in the $H_0$ value as a function of $n_T$. This difference with respect to the values preferred in the left panel is entirely driven by the addition of a prior on $H_0$ from local measurements. Clearly, the dataset base + HST allows higher values of $H_0$ when compared to only the base dataset, as well as when compared to the $\Lambda$CDM prediction from current CMB data, i.e., $H_0 = 66.93 \pm 0.62$ kms$^{-1}$ Mpc$^{-1}$ [37].

---

[4]In addition to the cosmological parameters, we also vary the nuisance foreground parameters [43].

[5]Inflation consistency is not applied to any of the models investigated in this work.

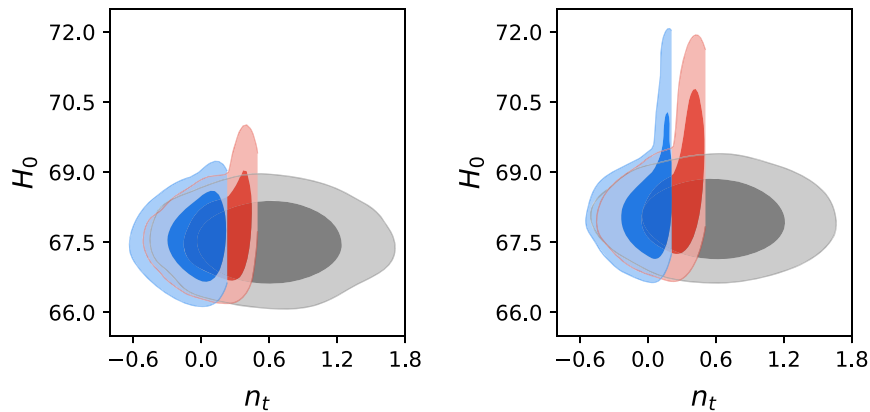

FIG. 3. Confidence regions (68% and 95%) for the plane $n_T - H_0$ of the Model 1 (blue), Model 2 (red), and $\Lambda$CDM with a relaxed consistency relation as the reference model (gray), using the base dataset (left) and base + HST (right).

In Table I we show the best fit values for the $H_0$ and $n_t$ parameters with 68% confidence limits obtained from the two analyzed datasets. We note that the $H_0$-tension problem is slightly alleviated by the contribution $N_{\rm eff}^{\rm GW}$. For the results obtained with the base dataset—which does not consider the $H_0 = 73.52 \pm 1.62$ measurement of Ref. [39]—we find that the tension is slightly reduced to $\sim 3.52\sigma$, $\sim 3.46\sigma$, and $\sim 3.46\sigma$, respectively, for the models $\Lambda$CDM $+ r + n_t$ and those with $k_{\rm UV} = 10^{22}$ (Model 1) and $k_{\rm UV} = 10^{54}$ (Model 2). The best fits and mean values are not in the limiting region corresponding to the steep increase of $H_0$ shown in Fig. 3, and only in this region can the tension be alleviated. In addition, the prior we consider for the optical depth from the latest Planck data [41] limits the value of $N_{\rm eff}^{\rm GW}$.[6] In spite of this, Fig. 3 clearly shows that a contribution of $N_{\rm eff}^{\rm GW}$ does have a role in alleviating the $H_0$-tension problem for models that predict $n_t$ in a certain range, i.e., $n_t \simeq \mathcal{O}(10^{-1})$, with the specific values depending on $k_{\rm UV}$.

As mentioned earlier, the GW's background energy density affects the expansion rate during the BBN. On the other hand, the BBN results impose limits on the value of $N_{\rm eff}$ which, from Eq. (8), implies constraints on $n_t$. Considering the cutoff choice $k_{\rm UV} = 10^{22}$ Mpc$^{-1}$ and a small value of the tensor to scalar ratio, e.g., $r \approx 10^{-3}$, we obtain from Eq. (8) that an upper limit of $N_{\rm eff} < 5$ implies $n_t < 0.5$, whereas by considering the more restrictive limit suggested in [51], $N_{\rm eff} < 3.56$, the constraint on $n_t$ changes only slightly to $n_t < 0.49$. On the other hand, in the same context, if we change the cutoff value to $k_{\rm UV} = 10^{54}$ Mpc$^{-1}$ we obtain $n_t < 0.2$, which is a very restrictive limit. Higher values of $r$ imply more restrictive constraints on $n_t$, in agreement with the results obtained in [31]. In Fig. 4 we show the impact of changing the values of $n_t$ and $H_0$. From this figure, it is clear that the $H_0$ tension can be alleviated or resolved in the context of the studied model only for positive values of $n_T$.

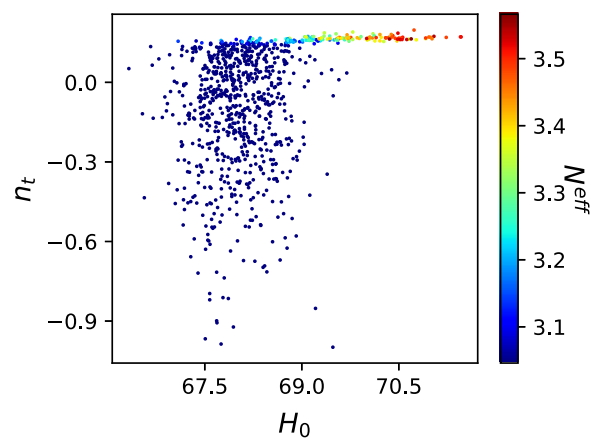

FIG. 4. Three-dimensional plot of $n_T - H_0 - N_{\rm eff}$ for the analysis using dataset base + HST and the model with $K_{\rm UV} = 10^{54}$ Mpc$^{-1}$.

Finally, it is worth mentioning that the BBN constraints on $n_t$ are particularly sensitive to the value of $k_{\rm UV}$. This result can be understood from the behavior shown in Fig. 1, which illustrates how $N_{\rm eff}$ varies with $k_{\rm UV}$ for some particular values of $n_T$. Unlike the constraints on $n_T$ from LIGO,[7] e.g., BBN imposes constraints on the quantity $N_{\rm eff}$, which involves an integral in $k$. In addition, as mentioned in [26], the BBN constraints depend on the value of both the primordial helium mass fraction and the baryonic density parameter.

---

[6]Previous CMB data allowed even higher values of $N_{\rm eff}$ and $H_0$ implying a stronger preference for models that predict an extra contribution to $N_{\rm eff}$ (see, e.g., [34,35]).

[7]Constraints from LIGO are more robust in the sense that they constrain directly $n_T$ instead of a quantity that can be degenerated to modifications on the neutrino sector. The present constraints from LIGO on $n_t$ establish the higher value of $n_t$ that would not imply a detectable value at LIGO—that was not yet measured. However, the scales at which LIGO is sensitive can be affected by astrophysical sources of GW stochastic backgrounds [52]. For this reason the limits that we obtain on $n_t$ can be regarded as conservative.

TABLE I. Confidence limits (68%) for the $H_0$ and $n_t$ parameters using the two datasets discussed in the text.

| | | $\Lambda$CDM + r + $n_t$ | $k_{UV} = 10^{22}$[a] Model 2 | $k_{UV} = 10^{54}$ Model 1 |
|---|---|---|---|---|
| Base | $H_0$[b] | $67.57 \pm 0.53$ | $67.55 \pm 0.56$ | $67.51 \pm 0.54$ |
| | $n_t$ | $0.46^{+0.36}_{-0.53}$ | $0.04^{+0.38}_{-0.14}$ | $-0.13^{+0.29}_{\pm0.11}$ |
| Base + HST | $H_0$ | $67.91 \pm 0.55$ | $68.49 \pm 0.83$ | $68.35 \pm 0.81$ |
| | $n_t$ | $0.45^{+0.51}_{-0.47}$ | $0.12^{+0.34}_{-0.13}$ | $-0.07^{+0.29}_{-0.10}$ |

[a]$k_{UV}$ is in units Mpc$^{-1}$.
[b]$H_0$ is in units kms$^{-1}$ Mpc$^{-1}$.

## V. CONCLUSION

A consistent analysis of the early universe should take into account the effects of primordial GWs on the background expansion. A higher contribution of the effective number of relativistic species, $N_{eff}$, at early times leads to a faster pre-recombination expansion rate and a smaller value of the sound horizon at recombination. The resulting reduction in the angle subtended by the CMB acoustic peaks can be compensated by an increase in the value of the current expansion rate, $H_0$ [33]. While extra neutrino species are not predicted by the standard particle physics, primordial GWs with a blue tensor spectrum can naturally provide a significant contribution for $N_{eff}$ and, therefore, alleviate the current $H_0$-tension problem [40].

In this paper we have considered the extra contribution from the primordial GWs' background to $N_{eff}$ and studied its consequences on the expansion history of the universe, particularly on the current constraints on the $H_0$ parameter. Considering three scenarios and two datasets involving recent CMB and BAO observations along with priors from LIGO and the HST measurements of $H_0$, we have shown that models with a blue tensor spectrum ($n_T > 0$) allow a higher value of $H_0$ than does the standard $\Lambda$CDM model (see Table I). Our results, therefore, show that for the models explored an additional contribution of $N_{eff}^{GW}$ to $N_{eff}$ does not solve the current $H_0$-tension problem but can alleviate it. Moreover, they reinforce the need to explore early universe models with a blue tensor spectrum, as predicted by some noninflationary scenarios. This is especially important nowadays considering the great improvement on the $N_{eff}$ constraints recently provided by the new polarization data from the Planck Collaboration and also the advances in constraining the tensor parameters expected from the next generation of experiments.

## ACKNOWLEDGMENTS

The authors thank Daniel Meerburg for the useful discussion. We also acknowledge the use of the High Performance Computing Center at the Universidade Federal do Rio Grande do Norte (NPAD/UFRN) for providing the computational facilities to run our analysis. L. G. and M. B. acknowledge financial support of the Fundação Carlos Chagas Filho de Amparo à Pesquisa do Estado do Rio de Janeiro (FAPERJ—fellowship *Nota 10*). L. G. also acknowledges support from Conselho Nacional de Desenvolvimento Científico e Tecnológico (CNPq) (301091/2017-0). M. B. is also supported by INFN, Napoli section, QGSKY project. J. A. acknowledges support from CNPq (Grants No. 310790/2014-0 and No. 400471/2014-0) and FAPERJ (Grant No. 233906). The authors also acknowledge the authors of the CosmoMC (A. Lewis) code.